\documentclass[10pt]{article}
\usepackage{amsbsy}
\usepackage{amssymb}
\usepackage{amsmath}
\usepackage{geometry}
\usepackage[british]{babel}

	\newcommand{\ncd}{\newcommand}
	\ncd{\mrm}    {\mathrm}
	\ncd{\beq} {\begin{equation}}
	\ncd{\eeq} {\end{equation}}

\begin{document}

\title{Heat conduction in relativistic systems: alternatives and perspectives\footnote{Invited article to the IoP Gravitational Physics Group newsletter, December 2010.}}
\author{C. S. L\'opez-Monsalvo\footnote{cslm1x07@soton.ac.uk}\\
{\it School of Mathematics, University of Southampton SO17 1BJ, UK}}

\date{\today}

\maketitle

\emph{Does a moving body appear cold?} This remarkably simple question, raised by the late Professor P. T. Landsberg some forty years ago, highlights a profound missing link in our current understanding of classical thermodynamics and relativity. Some might interpret this as the lack of a relativistic transformation law for temperature measurements (see \cite{landsberg,landsbergm,nakamura}). There is, however, another view independent of any possible answer. It  shifts the attention from a mere exercise in velocity transformations to an argument about thermal equilibrium between moving bodies and the inertial properties of heat \cite{israel, ehlers, carternotes, cartermiss}. Such view promotes the discussion about the missing Lorentz transformation for temperature to the more elaborate - and physically meaningful - problem of heat exchange in a relativistic setting. Here we present the essential structure of our recent efforts to tackle such problem from a variational approach \cite{heatpaper}.

 Let us begin our discussion of heat from a non-relativistic point of view, where the notion of causality plays a less fundamental role. The heat equation is obtained by assuming the validity of the first and second laws of thermodynamics together with Fourier's law relating  heat with temperature gradients in an \emph{un-relaxed} manner. This implies that  when two bodies at different temperatures are put  in thermal contact, heat spontaneously flows from the warmer to the colder without any delay.  Such conclusion cannot be satisfactory. One would expect the speed of thermal disturbances to be bound by the internal structure of the media they travel on. Cattaneo saw a way around this problem by introducing in a reasonable (but arbitrary) manner, a modification to Fourier's law which takes into account the characteristic time a material  takes to react to thermal stimuli \cite{cattaneo}
	\beq
	\label{fourier}
	{\bf q} = -\kappa \nabla T \quad \rightarrow  \quad \tau\dot{\bf q} + {\bf q} = -\kappa \nabla T.
	\eeq
Here ${\bf q}$ denotes the heat flux, $T$ the temperature, $\kappa$ the thermal conductivity, $\tau$ represents the relaxation time of the medium  and the dot denotes time differentiation. Such amendment leads to a telegrapher equation for the propagation of heat signals, with a finite cap on the speed they can reach. It is worth mentioning that Fourier's law is the simplest, but not the most general, possible ansatz to ensure that the second law is satisfied. This is done by explicitly making the change in entropy a quadratic function of the heat flow. We should keep this in mind in the forthcoming discussion.

In the relativistic case, the unbounded speed of thermal disturbances implied by the parabolic nature of the heat equation is more than mere inconvenience, it is indeed  a fundamental problem. To give context to our discussion, let us consider the case where  matter's motion is represented by a fluid whose normalised four velocity is given by $u^a$. The first attempt of a relativistic extension for the heat equation was due to Eckart \cite{eckartpr}. The kind of model he proposed has become a stereotype of a class of theories referred as \emph{first order}. In this class, the entropy current of the model, denoted by a vector field $s^a$ which is generally not aligned with the matter four-velocity, may only depend on terms which are linear in deviations from equilibrium, namely the heat flow or the shear viscosity.  For this class of theories, the simplest way to impose the second law, which locally takes the form $s^a_{\ ;a}\geq 0$, leads to a relativistic version of  Fourier's law
	\beq
	\label{eckart}
	q^a = - \kappa h^{ab}\left[T_{;b} + T \dot u_b \right], 
	\eeq
where $q^a$ represents the heat flux, $h^{ab}$ is a projector orthogonal to the matter flow and the semi-colon and dot denote covariant and proper time differentiation respectively. Being essentially identical to \eqref{fourier},  equation \eqref{eckart} inevitably produces  a non-causal theory. Furthermore, as shown by Hiscock and Lindblom \cite{hisc}, it also suffers from stability problems. However, Eckart's proposal exhibits an extra piece of information which is missing in the non-relativistic treatment; the acceleration term. This purely relativistic effect can naturally be interpreted as being due to an effective ``mass'' per unit entropy given precisely by the temperature (see discussions in \cite{ehlers,carternotes,heatpaper,nilsgreg}). It is worth  mentioning  that recently this term has been suggested to be the origin of the afore mentioned instabilities (see \cite{mexicans1,mexicans2}). However, here we adopt the vision that the appearance of the four acceleration in \eqref{eckart} is an inevitable, and physically meaningful, feature of any relativistic theory of dissipation.      

The failure of first order theories to produce a theory of heat conduction compatible with the principle of causality can be tracked down to their definition of the entropy current. In an effort driven by simplicity, it is not permissible to prematurely drop  higher order terms in deviations from equilibrium, since they may give rise to linear terms after the differentiation required by the second law. This point of view is at the heart of the class of \emph{second order theories} of heat conduction whose key contribution is the  widely known Israel \& Stewart model. Here, the entropy current used by Eckart is extended to include \emph{all} the possible second order combinations of dissipative effects \cite{Stewart77,Israndstew1,israndstew2}. This strategy, analogous to the Grad's 14-moment, is firmly ground on kinetic theory and provides a causal and stable account of relativistic dissipation.  It is not our intention here to further explain this particular framework, but to note that,  in spite of its success on stability and causality tests, the price to pay is the introduction of a set of second order couplings that, in principle, can be measured but which cannot be obtained within the realm of the theory. 

Owing to an increasing interest in hydrodynamic descriptions of high-energy relativistic plasmas, attention to relativistic theories of dissipation of the Israel \& Stewart type has been regained. In many such applications, whose main source of dissipation is due to viscosity, heat can effectively be relegated as a secondary effect. Indeed, it was this kind  scenario, in which the ratio of viscous to thermal resistivity can be pushed down to order unity, the main source of motivation for the development of Israel \& Stewart second order theory.  Our interests, however, follows from  quite a  different motivation. It ranges from recent efforts to model the dynamics of super-fluid neutron stars \cite{nilsgreg, nils2003} to high energy gases with photons providing the dominant pressure contribution \cite{carternotes}.

In this brief account of a long standing problem in relativity, it is our wish to shed some light upon the more  fundamentally  satisfactory \emph{variational} approach to relativistic heat conduction \cite{carternotes,cartermiss,carter1988}. Historically, due to an over simplified view of the components of the theory, this model, pioneered by Carter, failed the  tests of  stability and causality. However, it was later shown by Priou \cite{priou} that the predictions of Carter's \emph{complete} theory are equivalent, and at second order physically indistinguishable, to those of Israel \& Stewart.  Despite Priou's efforts, the issue has been closed in favour of the latter. 

Perhaps the most attractive feature of a variational construction of relativistic heat conduction is that, once the equation of state (or, equivalently, the Lagrangian density) of the system is known, the theory contains no free parameters. The price for this however is the inclusion of transport quantities in the ``Lagrangian'' density of the system. This is nevertheless inevitable in any circumstance where one's aspiration is to describe, at least to linear accuracy, situations departing from local thermal equilibrium.  

In Carter's variational approach to heat conduction, one considers a multi-fluid system whose species are represented by a particle number density current $n^a$ and an entropy flux $s^a$. These two currents, together with the spacetime metric, constitute the fundamental fields of the matter sector for the Einstein-Hilbert action. General covariance requires the Lagrangian to be a proper scalar, therefore, it should depend only on covariant combinations of its fundamental fields. If we consider the metric as a passive field, the Lagrangian density can only depend on combinations of the two fluxes, which includes the relative flow between them. This is precisely what we meant by the inclusion of transport quantities in the Lagrangian density. 

A constrained variation of the matter action, whose Lagrangian density has the characteristics described in the preceding paragraph and whose constrain is the one imposed by the conservation law of the  particle number density flux, allows us to write the local conservation of energy and momentum as a ``force balance'' equation.  It is worth noting that each of the individual ``forces'' appearing in such balance, takes a form completely analogous to the Lorentz force for electromagnetism in a general relativistic setting (see section 2 in \cite{heatpaper}). The requirement for the local energy conservation law to follow as a Noether identity of the variational principle,  only needs one of the currents to be strictly conserved, $n^a_{\ ;a}=0$ say.  This allows an ``extra'' freedom to allocate the second law of thermodynamics  by  taking into account the possibility in which the production term $s^a_{\ ;a}$ takes values different from zero. 

One of the central problems  to be faced before giving a \emph{real} thermodynamic interpretation to the variational construction, lies in the correct interpretation of temperature measurements. As stressed in the opening sentence, this is a non-trivial problem in relativity and, therefore, a choice of frame is forced upon us. In this simple case, where we have only one conserved current, the choice is somewhat natural. In the spirit of physical interpretation, we can chose to equip each infinitesimal part of the matter fluid with a thermometer and consider all physical measurements to be taken on, and with respect to, the matter frame. In such a case one can show that the projection of the  canonical conjugate momentum to the entropy flux into the matter frame, which we denote here by $\theta^\parallel$, corresponds to the thermodynamic temperature in the  Gibbs sense\footnote{The rate of change of energy with respect to entropy with all other independent thermodynamic quantities fixed.}. Having clarified this, it is not difficult to show that the equation for the heat flux relative to the matter frame takes the form \cite{heatpaper}
\beq
\label{heat}
\check{\tau}\left[\dot q^a + u^{c;a}q_c\right] + q^a = - \check{\kappa} h^{ab}\left[\theta^\parallel_{;b} + \theta^\parallel \dot u_b \right].
\eeq 
Here all the quantities, with the exception of $\check{\tau}$ and $\check{\kappa}$ which correspond to the \emph{effective} relaxation time and  thermal conductivity, have been properly introduced in the text. This result, which is a relativistic generalization of Cattaneo's equation [the modified Fourier's law \eqref{fourier}], follows directly from the variational principle together with the simplest assumption (in the sense previously discussed) to make the entropy production satisfy the second law of thermodynamics. 

The structure of \eqref{heat} combines all the features present in both \eqref{fourier} and \eqref{eckart}, including the acceleration term which in the variational context arises as a consequence of the equations of motion. The ``check'' marks  indicate that these quantities depend on deviations from thermal equilibrium which are higher than second order.  A comparison with an analogous expression obtained from the Israel \& Stewart theory  is beyond the scope of present means of verification.

Let us emphasize that the main objection to Carter's approach was based on issues about stability and causality. This is a consequence of the simplistic spirit of the original model, ignoring a crucial effect present in almost every multi-fluid system; \emph{entrainment}. In addition to Priou's argument on the equivalence up to second order of Carter and Israel \& Stewart theories,  using a recent generalization of the two-stream instability analysis \cite{twostream}, it is possible to assess the stability and causality of a proposed dissipative theory at the level of the equation of state or Lagrangian density. 
 
In conclusion, we have presented a summary of the most significant efforts  attempted to obtain a physically sound theory of heat conduction compatible with the principle of local causality. In spite of the recent stir made on first order theories, in our view, physical indicators such as the presence of second sound in materials hint that the correct physics to model departures from thermal equilibrium cannot be less than second order. Although the Israel \& Stewart model is the most prominent and widely used tool to describe dissipative systems, the additional couplings - necessarily introduced in their expansion - give the theory an effective, rather than fundamental, character. In this sense, the variational formalism not only provides us with an alternative to the Israel \& Stewart model, it  centres the attention in the dynamical actors of any canonical theory; the canonical conjugate momenta. This observation makes the multi-fluid construction a very powerful tool which may help us tackle deeper problems on non-equilibrium thermodynamics of relativistic systems.    

\vskip.35cm
{\small 
\noindent{\bf Acknowledgements}
I would like to express my gratitude to Nils Andersson for his sharp and insightful advice through my explorations on relativistic thermodynamics. Also, I am thankful to the Editor, Juan Valiente-Kroom, for the kind invitation to contribute to the present newsletter. Finally, I acknowledge Queen Mary College for its hospitality and CONACYT for financial support. 
}

\end{document}